\documentclass[namedreferences]{solarphysics}
\usepackage[optionalrh]{spr-sola-addons} 
\usepackage{graphicx}
\usepackage{epsFig}
\usepackage{color}                       
\usepackage{url}                         
\newcommand{\arcs}{\mbox{\ensuremath{^{\prime\prime}}}}

\begin{document}

\begin{article}

\begin{opening}

\title{Atmospheric Response of an Active Region to new Small Fux Emergence}

%
\author{D.~\surname{Shelton}$^{1}$\sep
       L.~\surname{Harra}$^{1}$\sep
       L.~\surname{Green}$^{1}$      
       }

%
\runningauthor{\textit{D.Shelton et al.}}
\runningtitle{Impact of Flux Emergence on Solar Atmosphere}

%
	\institute{$^{1}$ UCL-Mullard Space Science Laboratory, Holmbury St Mary, Dorking, Surrey, RH5 6NT, UK
                     email: \url{david.shelton.10@ucl.ac.uk} \\             
          }

\begin{abstract}
We investigate the atmospheric response to a small emerging flux region (EFR) that occurred in the positive polarity of Active Region 11236 on 23 \,--\ 24 June 2011. Data from the \textit{Solar Dynamics Observatory's Atmopheric Imaging Assembly} (AIA), the \textit{Helioseismic and Magnetic Imager} (HMI) and Hinode's \textit{EUV imaging spectrometer} (EIS) are used to determine the atmospheric response to new flux emerging into a pre-existing active region. 
 Brightenings are seen forming in the upper photosphere, chromosphere, and corona over the EFR's location whilst flux cancellation is observed in the photosphere. The impact of the flux emergence is far reaching, with new large-scale coronal loops forming up to 43 Mm from the EFR and coronal upflow enhancements of approximately 10 km s$^{-1}$ on the north side of the EFR. Jets are seen forming in the chromosphere and the corona over the emerging serpentine field. This is the first time that coronal jets have been seen over the serpentine field.
\end{abstract}

%

\end{opening}
\section{Introduction}
\indent Magnetic flux emergence describes the appearance of new flux at the solar surface and in the solar atmosphere.
This study looks at an emerging flux region (EFR) to determine how a small flux emergence event affects the pre-existing active region into which it emerges. In many cases, the magnetic flux emerges as a fragmented structure (e.g. \opencite{mt08}) called the serpentine field (first suggested by \opencite{ssttz96}). The serpentine field (for the rest of this study this will be referred to as ``the serpentine") is seen observationally as small-scale positive and negative polarities during the early stages of the emergence. The serpentine field is created by the interaction between the convective downflows and the emerging flux tube below the photosphere (\opencite{Cheungetal08}). The emerging flux tube rises through the convection zone until it loses buoyancy just below the photosphere. As magnetic field starts to pile up under the photosphere, convective upflows and downflows are able to change the shape of the magnetic field, creating undulations (\opencite{Cheungetal08}). This allows the $\Omega$-loops to emerge into the photosphere. However, the U-loops are not able to rise into the photosphere in the same way as the $\Omega$-loops due to the weight of the plasma on the U-loops draining from the $\Omega$-loop. Therefore the only way that the U-loop part of the undulated flux tube can fully emerge into the photosphere is by small-scale magnetic reconnection along the serpentine occurring at locations of current enhancements (\opencite{pasgrb04}). The serpentine was first discussed as a mechanism to explain the behaviour of U-loops and magnetic flux balance by \opencite{Spruitetal1987}. After successive reconnection events, the serpentine field assumes a global $\Omega$-loop configuration and interacts with the large-scale active region field around it (\opencite{hmhtow10}). In this way, the serpentine field can form part of the global active region by successive reconnections occuring between the $\Omega$-loops. If magnetic reconnection does facilitate the emergence of the serpentine field into the atmosphere, we should see some evidence for this in the form of brightenings, jets, new loops and upflow/downflow enhancements.


In the chromosphere, small-scale H$\alpha$ brightenings called Ellerman bombs (EB:\opencite{e17}) are thought to be an atmospheric response to successive magnetic reconnections in the serpentine field (\opencite{pasgrb04}). The Ellerman bombs have also been shown to heat the transition region (\opencite{srgdb04}). Observational evidence of the EFR interacting with the existing active regions is in the form of brightenings seen in other chromospheric lines such as Ca II H, which are produced by magnetic reconnection (\opencite{gzrbr08}). Emerging flux is also able to produce jets formed by magnetic reconnection between the EFR and the pre-existing magnetic structures. These jets have an inverted-Y shape, have lengths of up to 400 Mm and velocities of up to 300 km s$^{-1}$ (\opencite{shibataetal92}) and are seen in the X-ray.   These jets are also seen in the extreme ultraviolet in active region/EFR complexes (\opencite{gontikakisetal09}) with lengths up to 16 Mm and velocities up to 100 km s$^{-1}$ and EFR/polar coronal hole complexes (\opencite{khwmscw07}) with velocities of 30 km s$^{-1}$ and are seen at the \textbf{edge of the EFR}. Like the chromospheric jets, the jets in the corona also follow this inverted-Y shape (\opencite{mguu08}). These appear to be smaller versions of the X-ray jets seen in the corona (\opencite{shibataetal92}). 

Coronal responses to flux emergence include the formation of coronal loops, outflows, flares, and CMEs. New coronal loops are formed by magnetic reconnection between the magnetic field lines of the emerging flux and the pre-existing active region. 
In this article, we investigate what impact a small EFR has on a pre-existing active region and the local atmosphere. 
\textbf{In section \ref{Observations}, we discuss the instrumentation we used in this study and a brief history of the active region the EFR emerges into. In section \ref{Results}, we discuss our findings and then in section \ref{Discussion}, we discuss what our findings mean for the emergence of small-scale EFRs into pre-existing active regions. In section \ref{Conclusion}, we summarise our findings and conclusions.}

\section{Observations}
\label{Observations}
Active region 11236 was a northern hemisphere region and is located on the western limb during the time of this study (between 23 June 2011 and 25 June 2011). The active region had a negative leading polarity, which contained a sunspot, and a positive trailing polarity, which was dispersed. The active region first emerged on the far-side of the Sun on 5 June 2011 (as seen by STEREO-B). The active region had already been on the front-side of the Sun for approximately nine days before the new flux emerged and produced several flares $(\mathrm{C}$-class and below).  We used data from the AIA and HMI instruments onboard the Solar Dynamics Observatory and the EIS instrument onboard Hinode.\\
 The 193 \AA\/, 1600 \AA\/, and 304 \AA\ AIA wavelength channels were used at a 12 second cadence, 24 second cadence, and 60 second cadence respectively and the 45 second cadence line-of-sight HMI magnetograms were used. The AIA data have a pixel size of 0.6$\arcs$ and the HMI magnetograms have a pixel size of 0.6$\arcs$. All datasets were calibrated and co-aligned using the {\textsf aia\_prep} procedure in Solarsoft. This also holds for co-alignment between AIA and HMI. The HMI data were de-rotated to correct for area-foreshortening. In all magnetograms, the saturation threshold is set at $\pm$ 200G; this matches the average flux density that was measured for this EFR.
The 1600 \AA\ channel, the 304 \AA\ channel  and the 193 \AA\ channel from the AIA instrument are used to look for local intensity enhancements (brightenings) in the upper photosphere (5000 K), upper chromosphere (50,000 K) and the 1.25 MK corona and compare these brightenings to changes in the magnetic field. The brightenings are found using running difference images. Also, the 304 \AA\ and 193 \AA\ channels are used to look for EUV jets. In addition, the 193 \AA\ channel is used to look for any atmospheric response to the magnetic flux emergence away from the EFR. \\
The EIS data have an image temporal cadence of five minutes. The data coverage was 35 minutes per hour and the exposure time for each slit position was nine seconds. There is a gap of eight hours between 23 June 2011 at 23:00 UT and 24 June 2011 at 07:00 UT when EIS was not pointing at this active region. This EIS study used a sparse raster with a 2$\arcs$ slit and 4$\arcs$ step size. The EIS data were calibrated using {\textsf eis\_prep} and co-aligned with AIA by comparing features in the EIS 195 \AA\ intensity map and AIA 193 \AA\ data. The EIS intensity and Doppler velocity maps were created by fitting a single gaussian to the data. These maps are used to look for any upflow or downflow enhancements near the location of the flux emergence.

\section{Results}
\label{Results}
\subsection{Description of the Active Region Before Flux Emergence}

\begin{figure}[!ht]
\begin{center}
\includegraphics[width=0.95\textwidth]{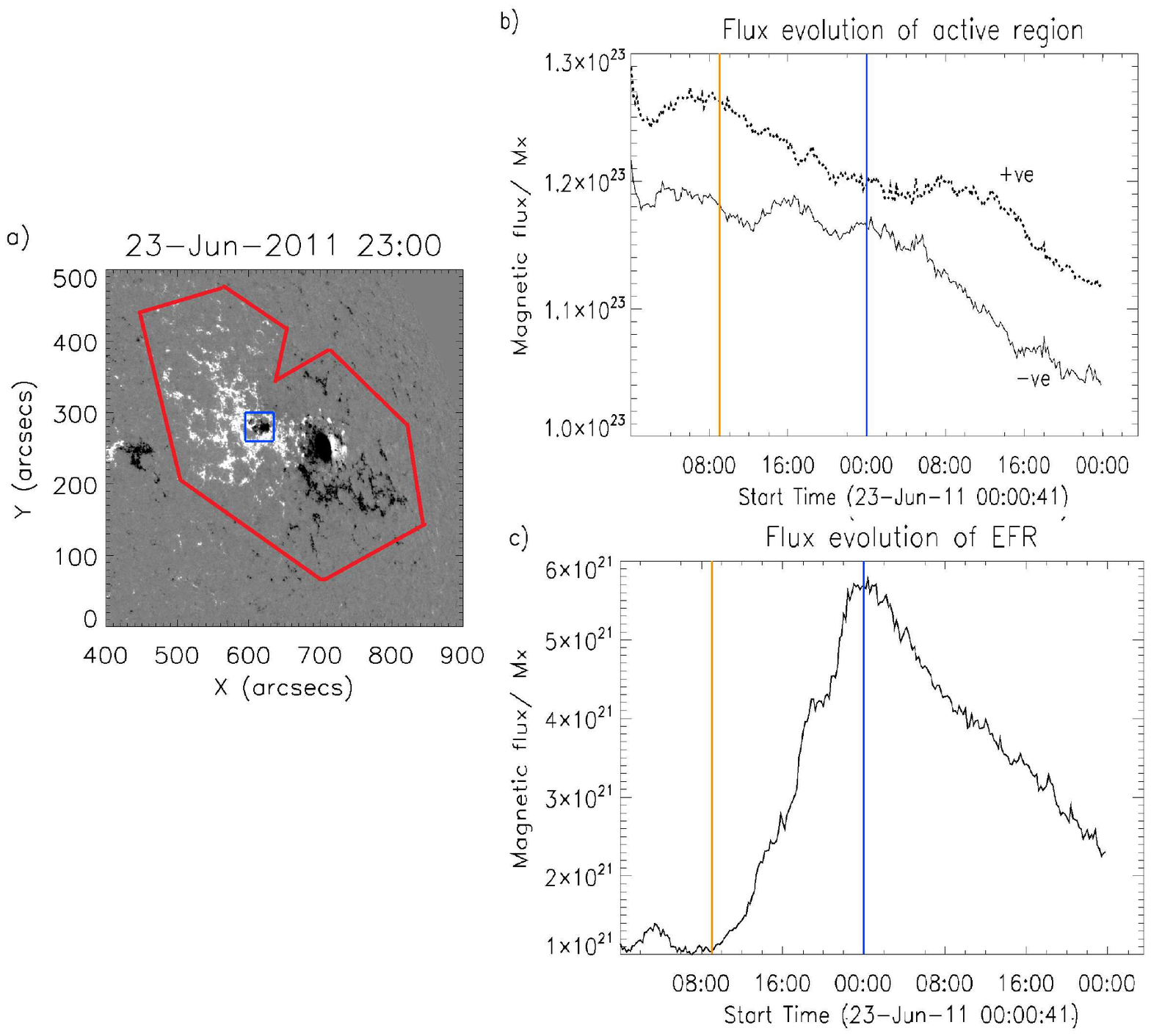}
\caption[flux evolution of EFR in AR11236] {(a) The area used to extract the flux evolution curves for the EFR (blue) and the active region (red). We used a minimum magnetic field strength of $\pm 25$ G. (b) The flux evolution of the active region; the dashed line is the positive flux and the solid line is the negative flux. (c) The flux evolution of the negative flux in the EFR. The two vertical lines in both flux evolution plots represent the start (first vertical line) of the flux emergence and the end (second vertical line) of the emergence phase of the EFR respectively. \textbf{The terms +ve and -ve refer to positive flux and negative flux.}}
\label{AR11236fluxcurves}
\end{center}
\end{figure}

 \begin{figure}[!ht]
\begin{center}
\includegraphics[width=0.8\textwidth]{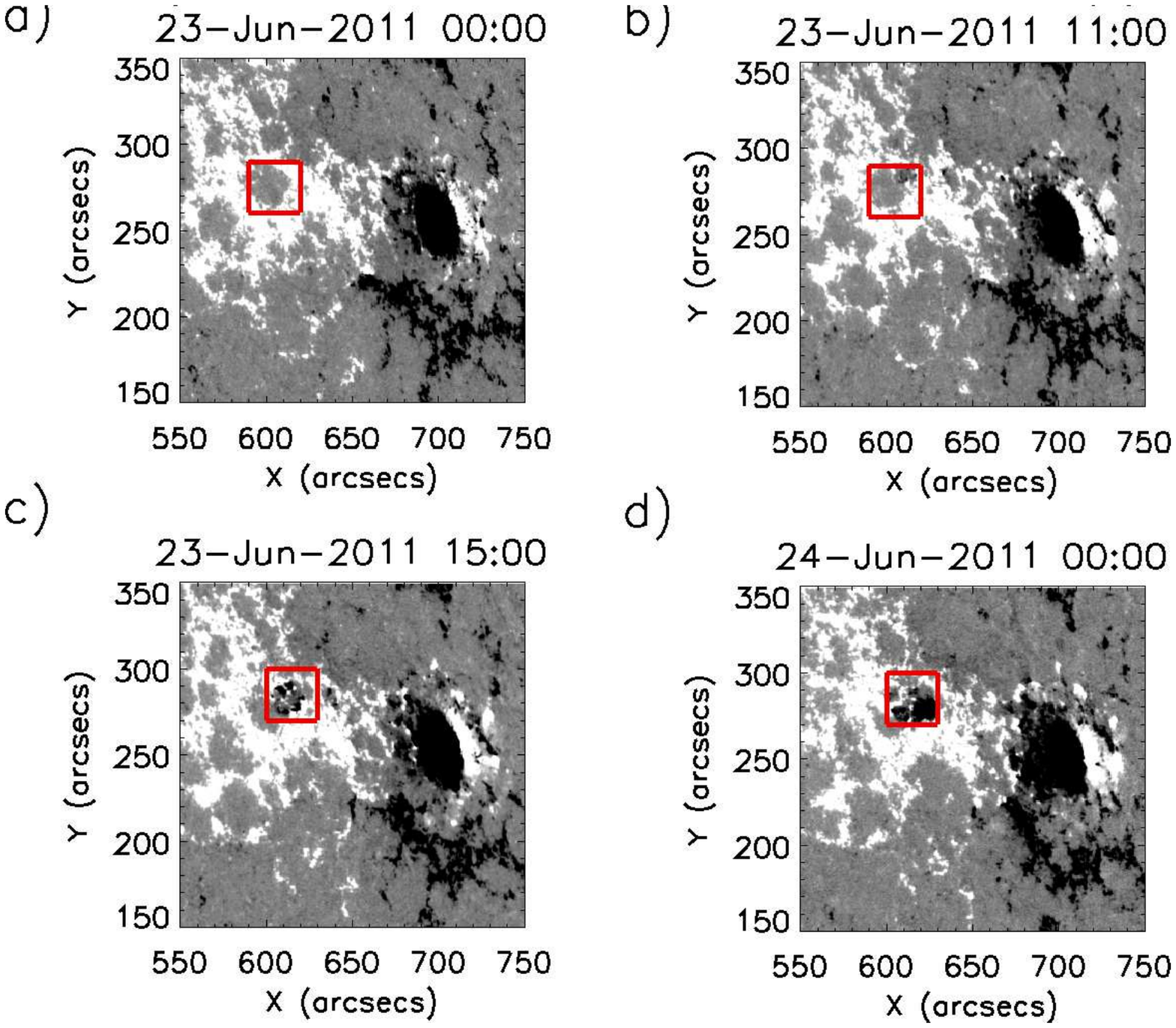}
\caption[HMI images of EFR evolution in AR11236]{The HMI magnetogram images show the changes seen in the active region by  before (a) and at \textbf{different stages of the flux emergence (FE)} (b, c and d). The red box shows where the new negative flux emerges.}
\label{11236hmievo}
\end{center}
\end{figure}

The active region is in decay with the magnetic flux over the two day period (Figure \ref{AR11236fluxcurves}b). We measured the active region flux by summing within the contour in Figure \ref{AR11236fluxcurves}a. This contour was drawn by eye and we used a minimum magnetic flux density of $\pm$ 25 G to remove the quiet Sun magnetic field from our findings. \opencite{closeetal2003} suggest that the quiet Sun magnetic flux density is 20 G. We see a difference between the amount of positive and negative flux. This is a line-of-sight issue and occurs when an active region is near the limb (in this case the western limb). \textbf{The flux trend has not been corrected for heliocentric angle variation as we are only interested in qualitiatively exploring how the flux changes in the active region and the emerging flux region. We are studying increases or decreases in magnetic flux over short periods of time (minutes).}
. The \textbf{first vertical line} in Figure \ref{AR11236fluxcurves} represents the start time of the flux emergence and the second vertical line represents the \textbf{end of the emergence phase of the} EFR. The flux emergence begins at 09:00 UT on 23 June 2011 in the photosphere and lasts for 15 hours (Figure \ref{AR11236fluxcurves}c). The negative magnetic flux of the EFR was measured by summing within a contour defined by the small box in Figure \ref{AR11236fluxcurves}a. We selected the box when the \textbf{EFR} had reached the end of the emergence phase. As the positive flux of the flux emergence region is imbedded with the positive flux from the pre-existing active region, we only look at the negative flux in the flux emergence region in Figure \ref{AR11236fluxcurves}c. Figure \ref{11236hmievo} shows different stages of the flux emergence: a, before the flux emergence, b, at the start of the flux emergence, c, during the flux emergence and d, at the end of the flux emergence. There is also a small emergence event that lasts for four hours between 01:00 UT and 03:00 UT on 23 June 2011. This then decays over the next two hours before disappearing from the magnetograms (Figure \ref{AR11236fluxcurves}c). \textbf{The small box in Figure \ref{AR11236fluxcurves}a is used} to measure the flux evolution to cover all the negative flux in the EFR.

\subsection{Atmospheric Response to Flux Emergence}

\begin{figure}[!ht]
\begin{center}
\includegraphics[width=0.70\textwidth]{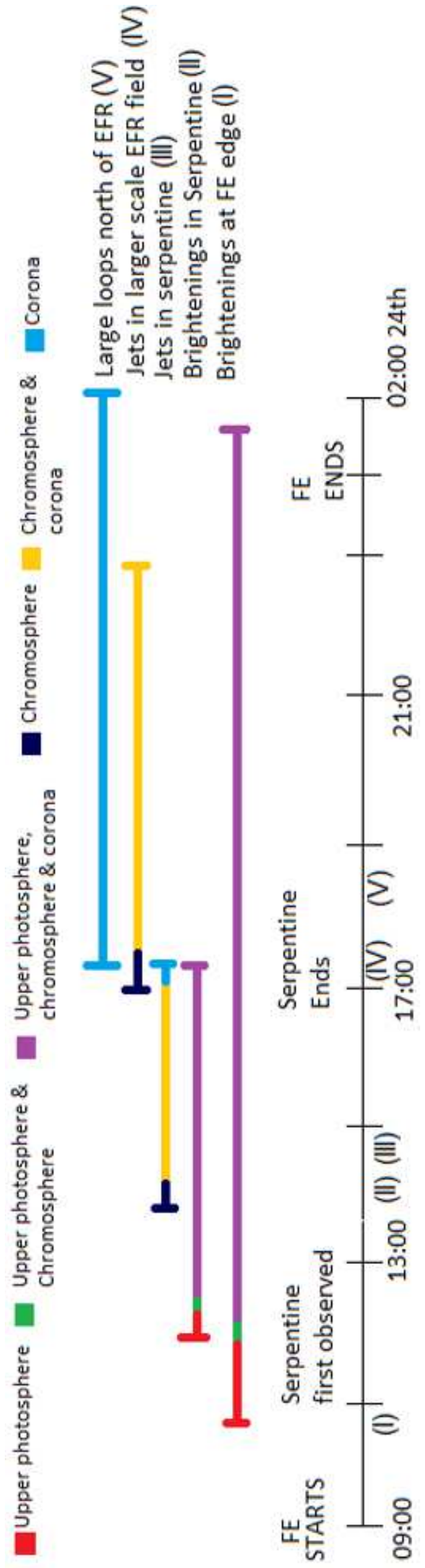}
\caption{Timeline showing events connected to the emergence of new flux in a pre-existing active region. The time periods are labelled as follows: brightenings seen at the \textbf{edge of the flux emergence (FE)}(I), brightenings seen in the serpentine (II), jets seen in the serpentine (III), jets seen over the large-scale EFR field (IV), new loops seen away from the flux emergence (V).}
\label{timelineEFR}
\end{center}
\end{figure}

We investigate how flux emergence is related to the events seen in the atmosphere on 23 June 2011. Brightenings started in the upper photosphere approximately two hours after the start of the flux emergence and continue until the end of the flux emergence in the upper photosphere, chromosphere and corona. These brightenings are seen at the \textbf{edge of the EFR} (this is referred to as time period I in Figure \ref{timelineEFR}). Brightenings are first seen in the serpentine field approximately 30 minutes after the serpentine was first observed and continues until the end of the serpentine emergence (this is referred to as time period II in Figure \ref{timelineEFR}). Chromospheric and coronal jets are first seen over the serpentine field approximately three hours (chromosphere) and 3.5 hours (corona) after the serpentine is first observed (this is referred to as time period III in Figure \ref{timelineEFR}). These jets last until the end of the serpentine emergence. The jets associated with the large-scale EFR field are seen in the chromosphere and corona after the serpentine field has fully emerged and are seen for approximately five hours (this is referred to as time period IV in Figure \ref{timelineEFR}). This is also the time when there is a rise in the number of large-scale coronal loops forming away from the EFR (this is referred to as time period V in Figure \ref{timelineEFR}). \textbf{The following sections show one example at each stage of the flux emergence highlighted in Figure \ref{timelineEFR}. We do not compare the examples to each other, and the aim is to show the atmospheric activity as a response to changes in the magnetic field. The changes in the magnetic field are not compared between examples and the blue box seen in the examples is determined by the size of the brightening in 1600 \AA\.}

\subsubsection{Brightenings at the Edge of the Flux Emergence}


\begin{figure}[!ht]
\includegraphics[width=1.00\textwidth]{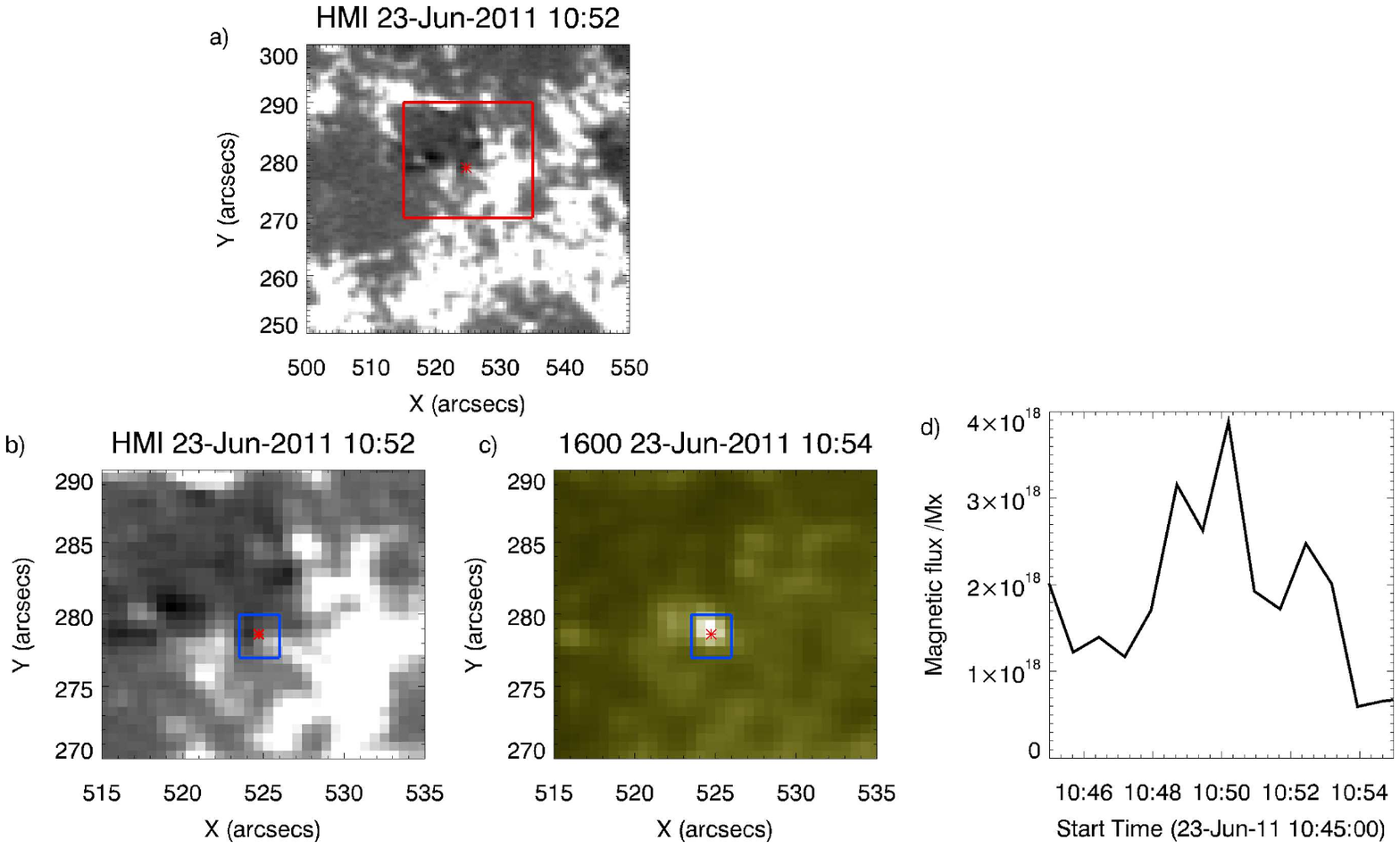}
\caption[First brightening seen in AR11236 EFR]{Images from HMI (a,b) and the 1600 \AA\ channel (c) on AIA when the brightening occurs in 1600. The red star represents the location of the brightening in the 1600 \AA\ channel. The large box in (a) shows the field of view of (b) and (c). The flux evolution profile for the negative polarity (d) was measured from the small box in (b).}
\label{brighteningstart}
\end{figure}
Brightenings are seen at the edge of the flux emergence region from approximately two hours after the flux emergence begins to the end of the emergence phase (as shown in Figure \ref{timelineEFR}, time period I).
The first brightening seen in the 1600 \AA\ (5000 K) occurs at 10:54 UT on 23 June 2011 (Figure \ref{brighteningstart}c) and is seen at the \textbf{edge of the EFR}. We attribute this brightening to a flux cancellation event as observed by HMI (as show in the electronic supplementary material as brighteningstart). At this location we see a small area of negative magnetic flux between two positive polarities (Figure \ref{brighteningstart}b). A decrease in flux is seen in Figure \ref{brighteningstart}d of approximately 2$\times10^{18}$ Mx and relates to when the cancellation is seen in the HMI movie.

\subsubsection{Brightenings in the Serpentine}

\begin{figure}[!ht]
\includegraphics[width=1.00\textwidth]{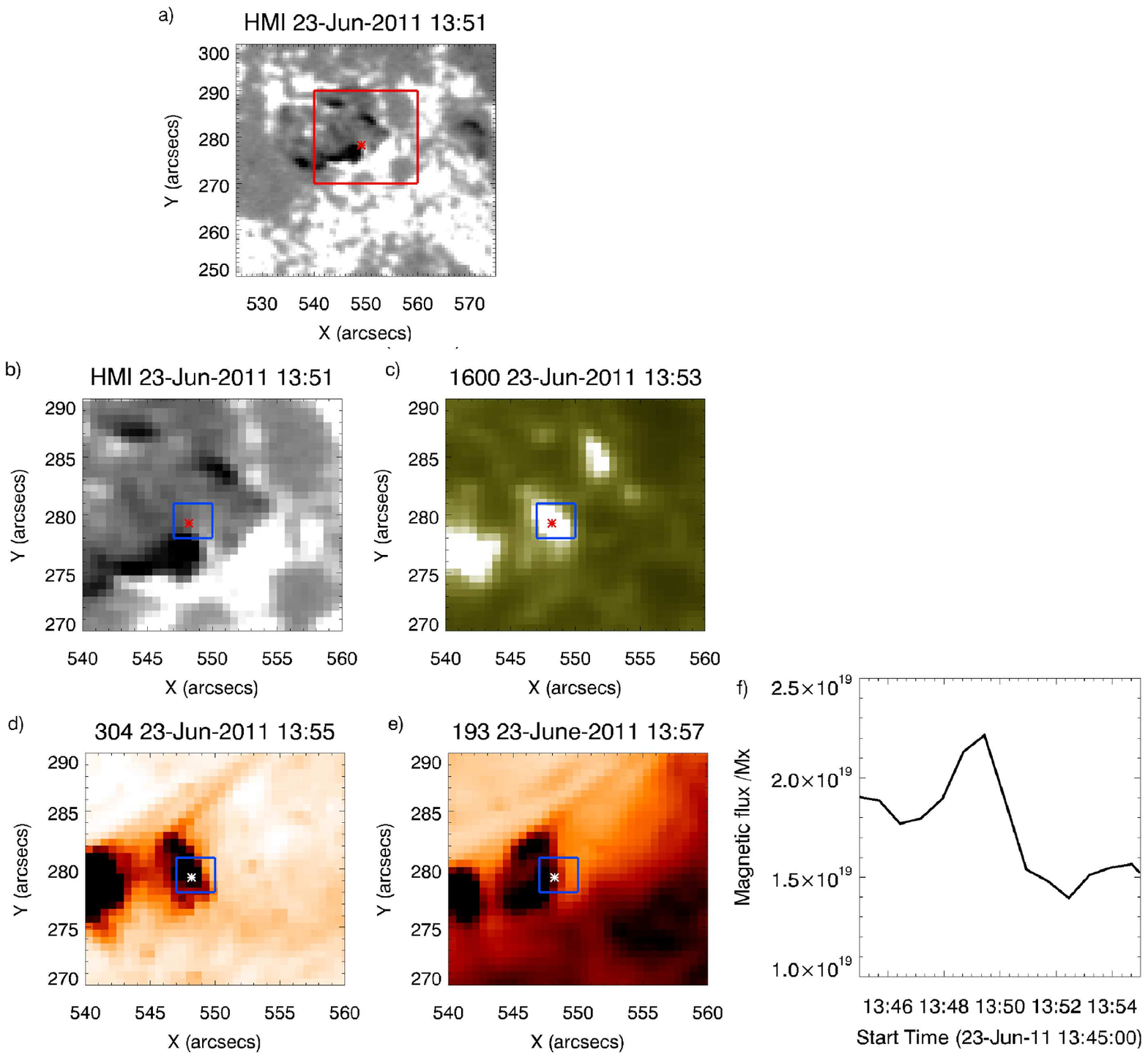}
\caption[Brightening seen at the edge of AR11236 EFR]{Images from HMI (a,b) and in 1600 \AA\ (c) , 304 \AA\ (d) and193 \AA\ (e) channels when a brightening occurs at the edge of the magnetic flux emergence region. The 193 \AA\ and 304 \AA\ images are shown in a reverse colour table. The star represents the location of the brightening in the 1600 \AA\ channel. The box in panel (a) shows the field of view of panels (b) to (e).  The flux evolution profile of the negative polarity (f) was measured from the small box in (b).}
\label{brightening1}
\end{figure}


The serpentine emerges between 11:10 and 17:00 on 23 June 2011.
Brightenings are seen in the serpentine from approximately 30 minutes after the serpentine starts to emerge to the end of the serpentine emergence (as highlighted in Figure \ref{timelineEFR}, time period II).
One example of brightenings in the serpentine field  is at 13:51 UT (Figure \ref{brightening1}a). We see this brightening in 1600 \AA\ (upper photosphere), in 304 \AA\ (upper chromosphere) and 193 \AA\ (corona at 1.25 MK). We see this brightening at 13:53 UT (Figure \ref{brightening1}c), 13:55 UT (Figure \ref{brightening1}d) and 13:57 UT (Figure \ref{brightening1}e) respectively. This brightening is associated with magnetic flux cancellation between the major negative polarity and the positive polarity of a small emerging bipole that occurs at 13:49 (Figure \ref{brightening1}b). \textbf{A decrease in flux is seen} in Figure \ref{brightening1}f of approximately 1$\times10^{19}$ Mx and relates to when the cancellation is seen in HMI (as shown in the movie entitled brightening1). The brightening in 1600 \AA\ lasts for four minutes, the brightening in 304 \AA\ lasts for approximately four minutes, while the brightening seen in 193 \AA\ lasts for approximately three minutes. 

\subsubsection{Jets in the Serpentine}

\begin{figure}[!ht]
\begin{center}
\includegraphics[width=0.95\textwidth]{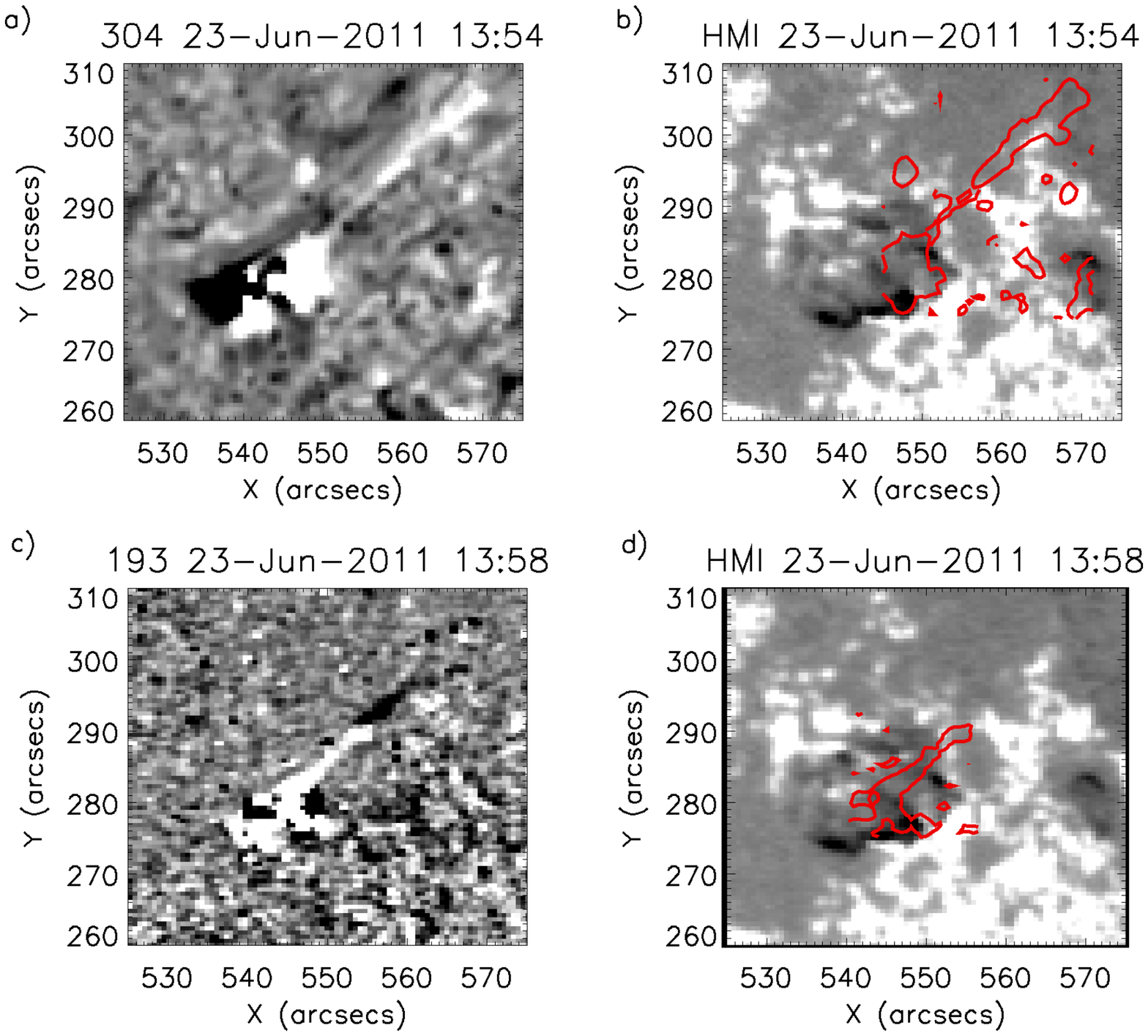}
\caption[First chromospheric jets seen in AR11236 EFR]{The first jets seen in the chromosphere (304 \AA\/) and corona (193 \AA\/) five hours after the flux emergence begins. (a): A running difference AIA 304  \AA\ intensity map of the jet between 13:48 and 13:54 (the image from 13:48 is removed from the image at 13:54) on 23 June 2011. (b): The closest HMI image to when the 304 \AA\ jet is released. The red contours in (a) and (b) show the location of the jet seen in the 304 \AA\ difference image. (c): A running difference AIA 193  \AA\ intensity map of the jet between 13:58 and 14:02 (the image from 13:58 is removed from the image at 14:02) on 23 June 2011. (d): The closest HMI image to when the 193 \AA\ jet is released. The red contours in (c) and (d) show the location of the jet seen in the 193 \AA\ difference image.}
\label{firstjets}
\end{center}
\end{figure}

The first chromospheric jets are seen in the AIA 304 \AA\ wavelength channel approximately five hours after the flux emergence begins (Figure \ref{firstjets}a) . The chromospheric jet velocities range between 20 and 30 km \textbf{s$^{-1}$} for the period of time jets are seen (approximately seven hours). We see the chromospheric jets over the serpentine field as well as at the edges. 

\begin{figure}[!ht]
\includegraphics[width=1.00\textwidth]{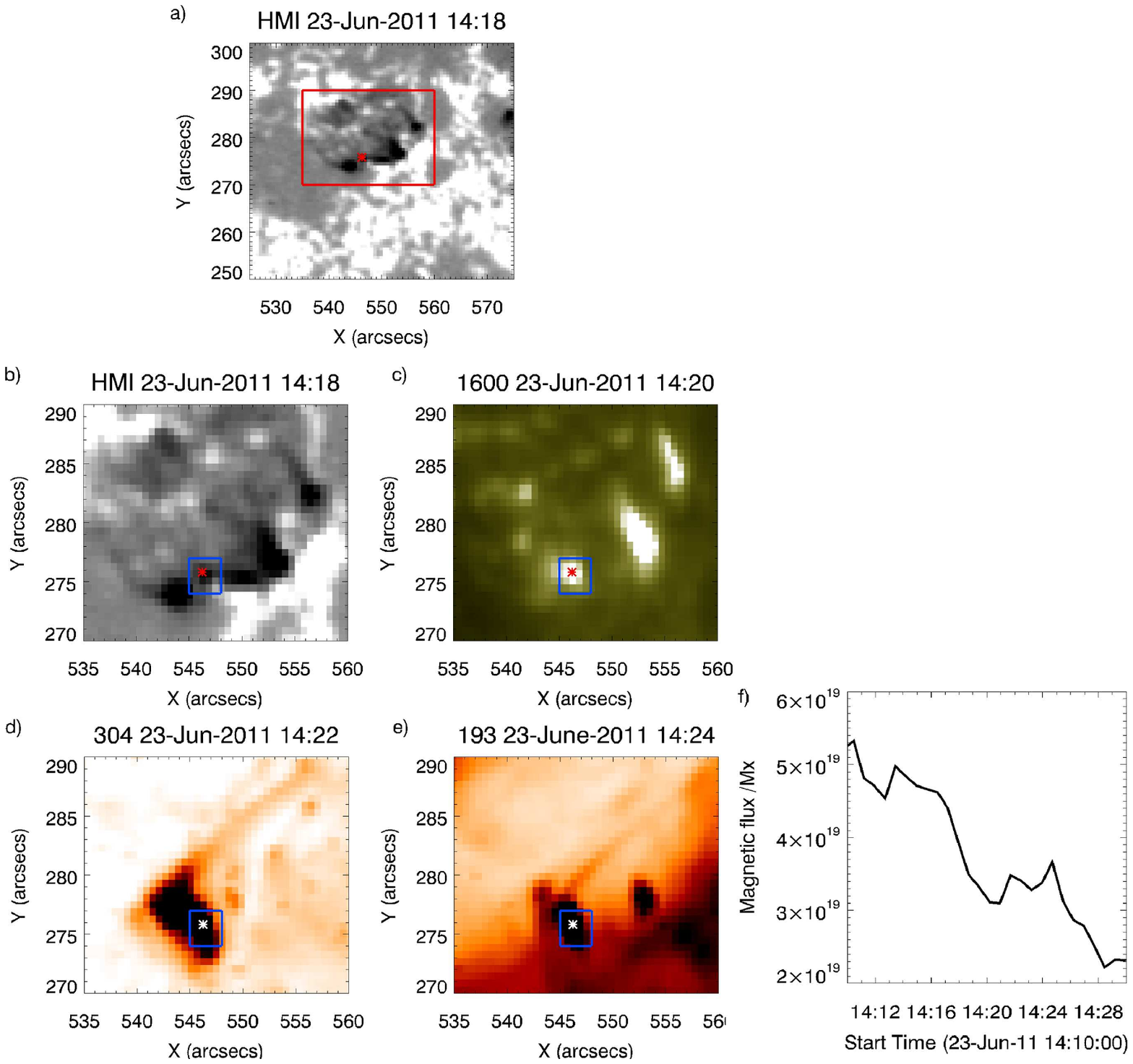}
\caption[Brightening associated with jet in AR11236 EFR]{Images from HMI (a,b) and in 1600 \AA\ (c), 304 \AA\ (d) and 193 \AA\ (e) channels when a brightening occurs at the same time that chromospheric and coronal jets are formed over the serpentine at approximately 14:15. The 193 \AA\ and 304 \AA\ images are shown in a reverse colour table. The star represents the location of the brightening in the 1600 \AA\ channel. The box in panel (a) shows the field of view of images (b) to (e).  The 193 \AA\ and 304 \AA\ panels are shown with a reverse colour table. The flux evolution profile for the positive polarity (f) was measured from the small box in (b).}
\label{brightening4}
\end{figure}

The first coronal jets are seen in the AIA 193 \AA\ wavelength channel approximately four minutes after the first chromospheric jets. There are no coronal jets seen over this region before the flux emergence begins. At the early stages of the jet formation (for example, Figure \ref{firstjets}c), these jets have a velocity of approximately 50-60 km \textbf{s$^{-1}$}. They are spatially located over the region of magnetic flux emergence with the footpoints appearing to be in the serpentine field rather than at the major polarities of the EFR. \textbf{This is the first time that coronal jets have ever been seen over any serpentine}. The chromospheric and coronal jets are seen over the serpentine from approximately two hours after the start of the serpentine emergence until the end of the serpentine emergence (as highlighted by Figure \ref{timelineEFR}, time period III).

There are brightenings in the 1600 \AA\ channel that are co-spatial with the jets seen in the 304 \AA\ and 193 \AA\ channels. \textbf{An example that shows that the effects of flux emergence} occurs at 14:18 UT in HMI (Figure \ref{brightening4}a), in 1600 \AA\ at 14:20 UT (Figure \ref{brightening4}c), in 304 \AA\ at 14:22 UT (Figure \ref{brightening4}d) and in 193 \AA\ at 14:24 UT (Figure \ref{brightening4}e). The brightening is associated with flux cancellation between a very small bipole and a larger negative polarity (Figure \ref{brightening4}b). \textbf{A decrease in flux is seen} in Figure \ref{brightening4}f of approximately 2$\times10^{19}$ Mx and relates to when the cancellation is seen in HMI (as shown by the electronic supplementary material entitled brightening2). This demonstrates the temporal and spatial connection between the brightenings observed in 1600 \AA\ at 14:20 UT and the jets seen in 304 \AA\ and 193 \AA\ at 14:22 UT and 14:24 UT. The  brightening in 1600 \AA\ lasts for eight minutes and the jets in 304 \AA\ and 193 \AA\ last for approximately five minutes.

\subsubsection{Jets in the Large-Scale EFR Field}


The later stage of the magnetic flux emergence is defined as occuring after the serpentine field is no longer observed. This occurs after 16:45 UT on 23 June 2011 and continues until approximately 01:00 UT on 24 June 2011. Chromospheric and coronal jets are seen over the large-scale field between the approximate end of the serpentine emergence to two hours before the end of the emergence phase (as highlighted by Figure \ref{timelineEFR}).

\begin{figure}[!ht]
\begin{center}
\includegraphics[width=1.00\textwidth]{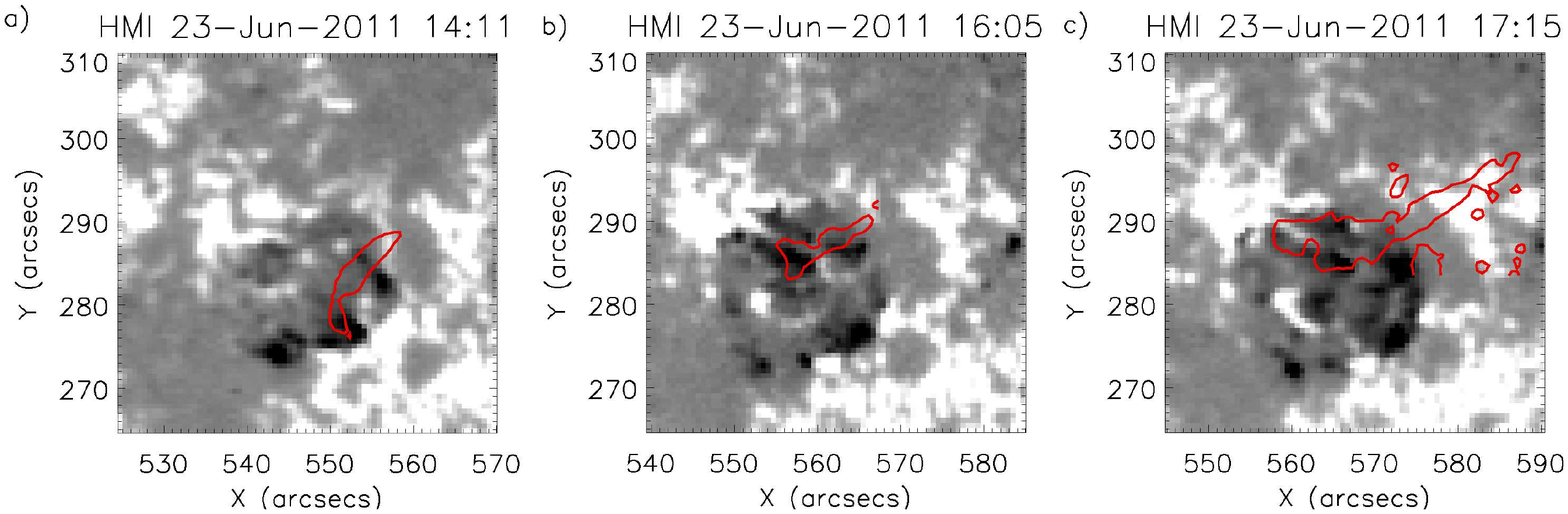}
\caption[Jets seen over main flux emergence field]{Jets seen migrating across the serpentine field. The contours show the jets. The jets start at the boundary between the flux emergence and the pre-existing active region (a), then move to the centre of the flux emergence (b) and then move to the north boundary between the flux emergence and the pre-existing active region (c).}
\label{jetlate}
\end{center}
\end{figure}

\begin{figure}[!ht]
\includegraphics[width=1.00\textwidth]{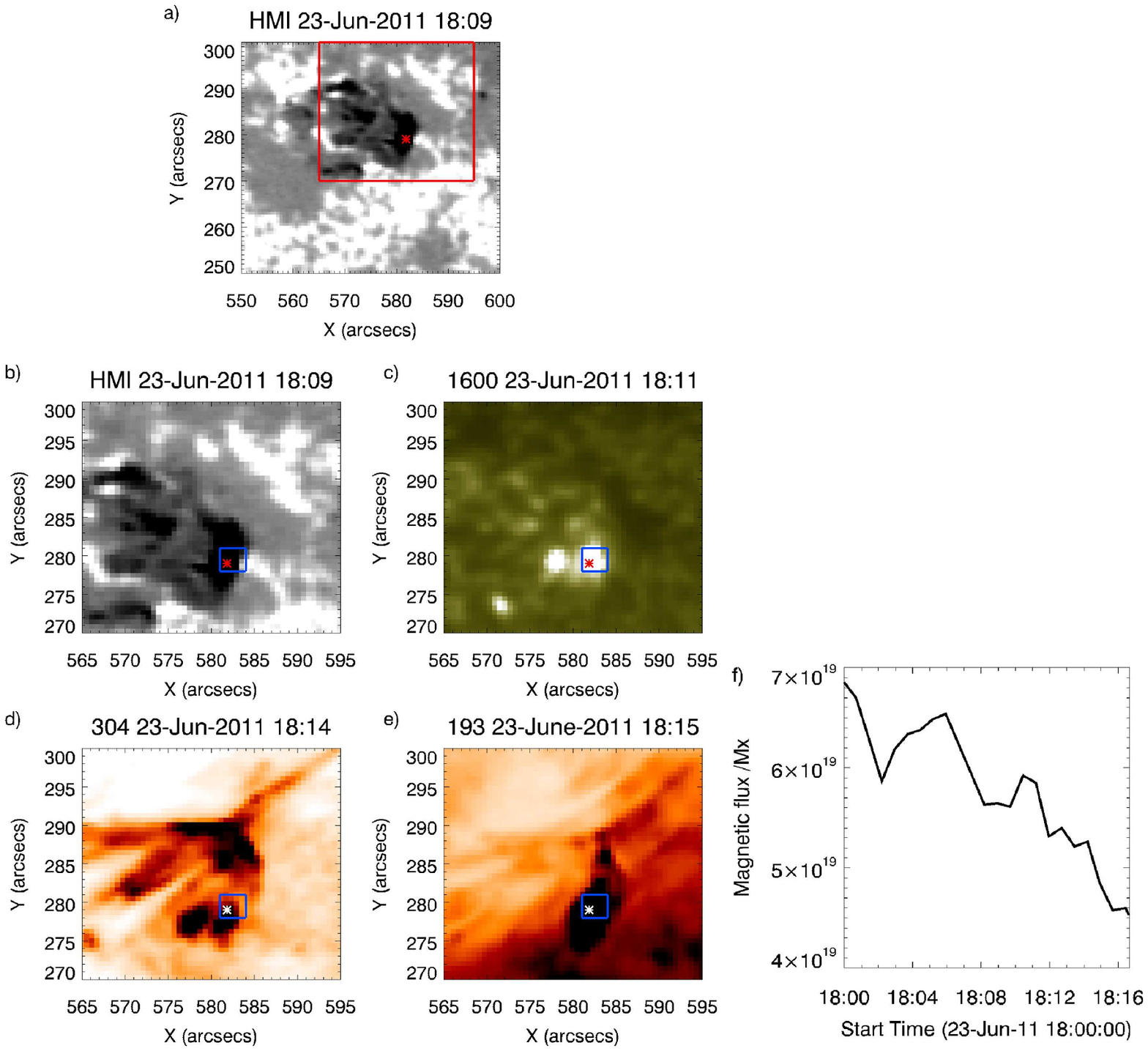}
\caption[Brightening associated with a jet formed over large-scale EFR field] {Images from HMI (a,b) and in 1600 \AA\ (c) , 304 \AA\ (d) and 193 \AA\ (e) channels when a brightening occurs at the same time chromospheric and coronal jets are formed over the large-scale EFR field at approximately 18:15.  The star represents the location of the brightening in the 1600 \AA\ channel. The large box in image (a) shows the field of view of images (b) to (e).  The 193 \AA\ and 304 \AA\ images are shown with a reverse colour table. The flux evolution profile for the positive polarity (f) was measured from the small box in (b).}
\label{brightening5}
\end{figure}

As the serpentine emerges, the jets appear to move across the flux emergence towards the north part of the EFR (Figure \ref{jetlate}). The jets start at the boundary between the flux emergence and the pre-existing active region (Figure \ref{jetlate}a) , then start forming over the serpentine field (Figure \ref{jetlate}b), before forming over the boundary between the flux emergence and pre-existing active region in the north (Figure \ref{jetlate}c). After this, the serpentine field has fully emerged and jets form between the large-scale EFR field and the pre-existing active region. These jets form at the EFR's south boundary with the pre-existing active region and last for 5 hours (as highlighted by Figure \ref{timelineEFR}, time period IV).

Brightenings are also associated with jets over the larger EFR field.
\textbf{An example that shows the effects of the larger EFR field occurs} at 18:08 UT in HMI (Figure \ref{brightening5}a), in 1600 \AA\ at 18:10 UT (Figure \ref{brightening5}c), in 304 \AA\ at 18:14 UT (Figure \ref{brightening5}d) and in 193 \AA\ at 18:15 UT (Figure \ref{brightening5}e). The brightening is associated with magnetic flux cancellation between the major negative polarity of the flux emergence and the pre-existing positive polarity of the active region (Figure \ref{brightening5}b). A decrease in flux is seen in Figure \ref{brightening5}f of approximately 1.5$\times10^{19}$ Mx and relates to when the cancellation is seen in HMI (as shown by the electronic supplementary material entitled brightening3). This demonstrates the temporal and spatial connection between the brightenings observed in 1600 \AA\ at 18:10 UT and the jets seen in 304 \AA\ and 193 \AA\ at 18:14 UT and 18:15 UT. The brightening seen in 1600 \AA\ lasts for six minutes and the jets seen in 304 \AA\ and 193 \AA\ last for approximately seven to ten minutes.

\subsubsection{Formation of New Loops Away From the EFR}
Up until now, we have investigated the atmospheric response to the emergence of the serpentine field and the interaction between the EFR and the positive polarity of the pre-existing active region. But is there any atmospheric response as we move away from the EFR and if so, how far away does the EFR's influence extend?

\begin{figure}[!ht]
\begin{center}
\includegraphics[width=0.65\textwidth]{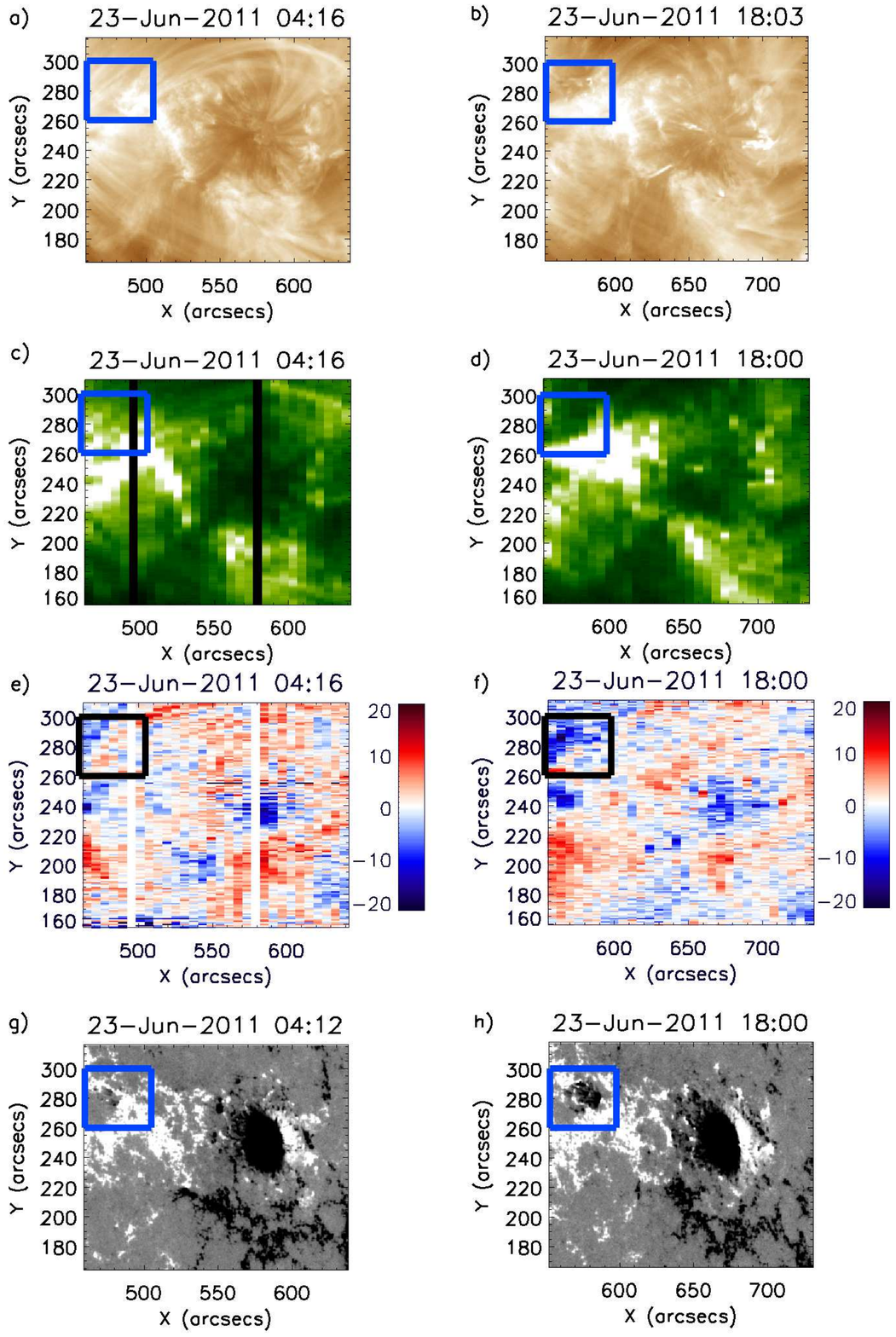}
\caption[Images showing upflow enhancements over EFR] {(a) and (b): AIA 193 \AA\ intensity maps before (five hours) and after (nine hours) the flux emergence begins. (c) and (d): Shows EIS Fe XII intensity maps before (five hours) and after (nine hours) the flux emergence begins. (e) and (f): EIS Fe XII Doppler velocity maps before (five hours) and after (nine hours) the flux emergence begins. The Doppler velocity range is between $\pm 20$ km $s^{-1}$. (g) and (h): two HMI magnetogram maps before (five hours) and after (nine hours) the flux emergence begins. The minimum magnetic field strength is $\pm 25$ G. The boxes show where the new flux region will emerge.}
\label{EIShmi}
\end{center}
\end{figure}

In the EIS velocity maps, blueshift enhancements are seen after the magnetic flux emergence begins (Figure \ref{EIShmi}f), and there is also an intensity enhancement in the EIS intensity maps after the magnetic flux emergence begins (Figure \ref{EIShmi}b). These enhancements were not seen before the magnetic flux emergence (Figure \ref{EIShmi}g). The blueshifts are located where positive and negative flux meet at the north part of the EFR (Figure \ref{EIShmi}h) and where there are intensity enhancements in the EIS 195 \AA\ data (Figure \ref{EIShmi}d). 

\begin{figure}[!t]
\begin{center}
\includegraphics[width=0.50\textwidth]{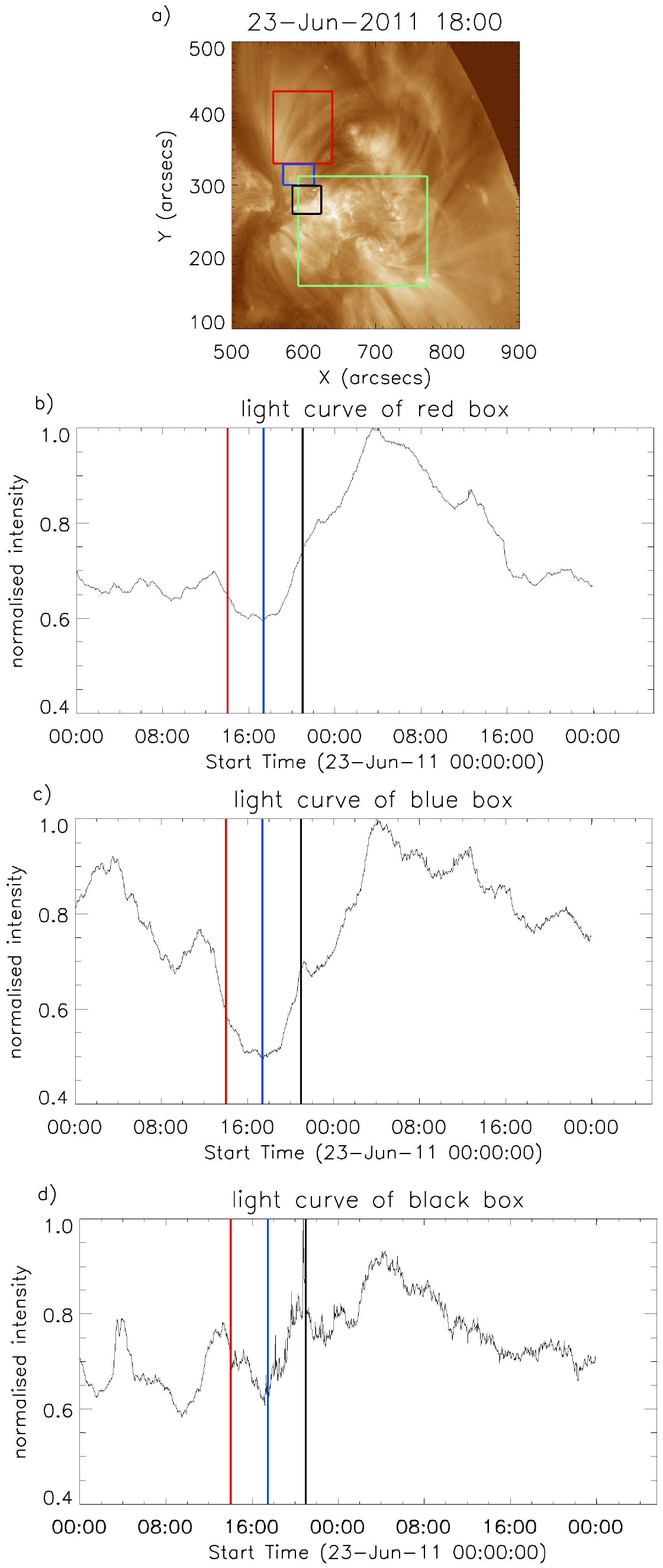}
\caption[light curves showing enhancement in intensity to the North of EFR] {(a): An AIA 193 \AA\ image with the light curve extraction regions overlaid. We measure the intensity evolution in the loops to the North (top and middle box) and around the EFR (small black box). The large bottom box shows the EIS field of view. (b),(c), and (d): the light curves for the top box, small middle box and small black box are shown in panel (a). The first vertical line shows the start time of the first jets seen at approximately 14:00 UT, the second vertical line shows the start of the intensity rise seen after the start of the jets, and the third vertical line shows the end time for jets. This rise is determined as a continual increase in intensity over a extended period of time. The second panel shows the light curve for the top box, the third panel shows the light curve for the small, middle box and the bottom panel shows the light curve for the small black box. The light curves for the small middle box and small black box cover the north-eastern part of the EIS field of view.}
\label{AIAlightcurves}
\end{center}
\end{figure}

 The velocities of the box in Figure \ref{EIShmi}e (before the magnetic flux emergence) and the velocities of the blue (black) box in Figure \ref{EIShmi}f (during the magnetic flux emergence) were compared and we find that the velocities in Figure \ref{EIShmi}f had an average upflow velocity enhancement of 10 $\pm 4$ km s$^{-1}$ when compared with the velocities in Figure \ref{EIShmi}e.

\begin{figure}[!ht]
\begin{center}
\includegraphics[width=0.80\textwidth]{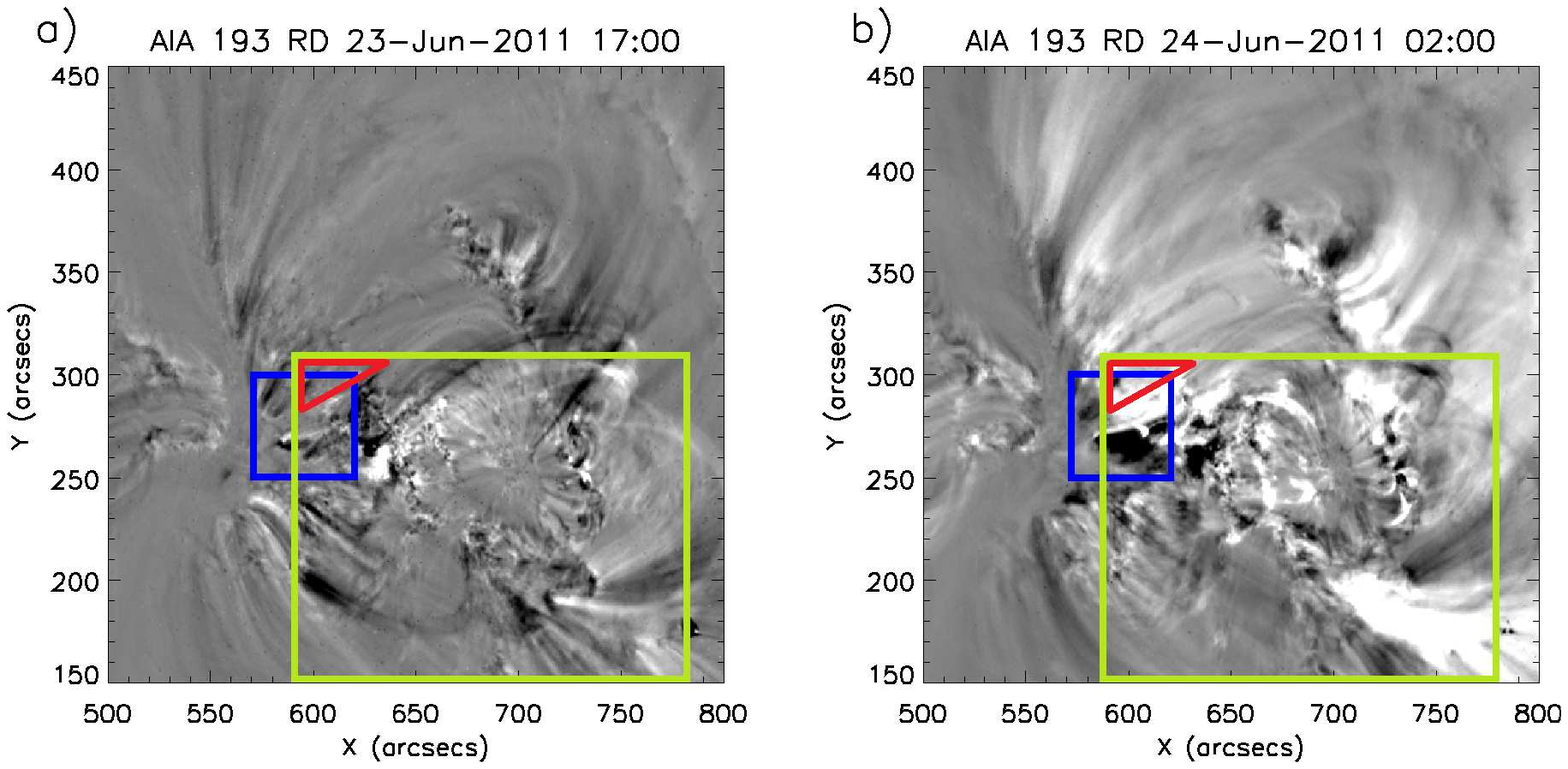}
\caption[Running difference showing new loop formation]{Running difference images before the rise in intensity (a) and at the peak of the intensity rise (b). We can clearly see an enhancement in intensity to the north of the EFR (highlighted by the small box). The large box represents the EIS field of view and the triangle represents the area in which we see the upflow enhancements.}
\label{loopproof}
\end{center}
\end{figure}

\indent In Figure \ref{AIAlightcurves}a, we show light curves that were extracted for three of the four boxes shown. \textbf{These three boxes cover: the EFR (small, black), the coronal loops north of the EFR (small, middle) and the coronal loops north of the blue box (top)}. The top box is approximately 100\arcs\ (43 Mm) away from the EFR. The EIS field-of-view is highlighted by the large, bottom box. These regions are used to try and understand the extent of the EFR's effect on loops that are far away from the EFR. In the intensity plots of these regions (Figure \ref{AIAlightcurves}b to d), \textbf{the start of the jets in the EFR is represented by the first vertical line. There is a rise in 193 \AA\ intensity in all three light curves that starts approximately three hours after the jets begin (as shown by the second vertical line) and ends just after the jets end (as shown by the third vertical line)}. This intensity rise is attributed to new plasma appearing in our extraction region. Therefore, this rise in intensity is expected to be associated with new loops forming over the extraction regions, and for one of the footpoints of these new loops to be located to the North of the EFR (as seen in Figure \ref{loopproof}). The small box shows the location of the EFR at 18:00 UT on 23 June 2011. It can be seen that the enhanced blueshifts are located in the northern part of the EFR (as shown by the red triangle in Figure \ref{loopproof}) which is associated with the lower legs of the large-scale loops. This suggests that the cause of the blueshifts is chromospheric evaporation. These large-scale loops continue to exist after the emergence phase has ended (this is highlighted in Figure \ref{timelineEFR}, time period V).

\section{Discussion}
\label{Discussion}
\subsection{The Early Stage of the Flux Emergence}
\indent  The first brightenings seen during the early stage of the flux emergence (Figure \ref{brighteningstart}) \textbf{are likely due to reconnection} occuring between the flux emergence and the pre-existing field. This area is a favourable site for magnetic reconnection because we see positive and negative polarities interacting with each other and a decrease in flux where the polarities are interacting. As a result of magnetic reconnection, one of the new bipoles is seen disappearing and a brightening occurs above this location (as shown in Figure \ref{brighteningstart}d). In previous studies, brightenings in the atmosphere are one of many observational pieces of evidence for magnetic reconnection (e.g. \opencite{guglielminoetal10}). Magnetic flux cancellation (where a bipole ``disappears") is also another indicator of magnetic reconnection.  If reconnection occurs in the lower parts of the atmosphere, one loop could have a small enough \textbf{radius of curvature so that} magnetic tension can pull the loop under the photosphere (\opencite{vanballegooijen89}). \textbf{In magnetograms, this is described magnetic flux retraction (e.g. \opencite{vdg2002}). \opencite{Kuboetal2010} suggests that flux cancellation events are highly time dependent and occur on scales \textless 200 km.} 

\subsection{The Emergence of the Serpentine Field}
The brightenings in the serpentine field (Figure \ref{brightening1}) are a response to magnetic reconnection and follows the model of \opencite{pasgrb04}. In the model of \opencite{pasgrb04}, brightenings in the wings of the H$\alpha$ line (called Ellerman bombs) are located over flux cancellation sites. The magnetic flux cancellation sites are favourable locations for magnetic reconnection and successive reconnection events can build up longer magnetic \textbf{field lines} in the serpentine field. Observations for the emergence of U-loops would require two $\Omega$-loops to be present, close to each other and emerge at the same time (\opencite{vdg2002}). \opencite{Magara2011} suggests that only shallow U-loops can emerge and cause flux cancellation.

\begin{figure}[!ht]
\includegraphics[width=1.00\textwidth]{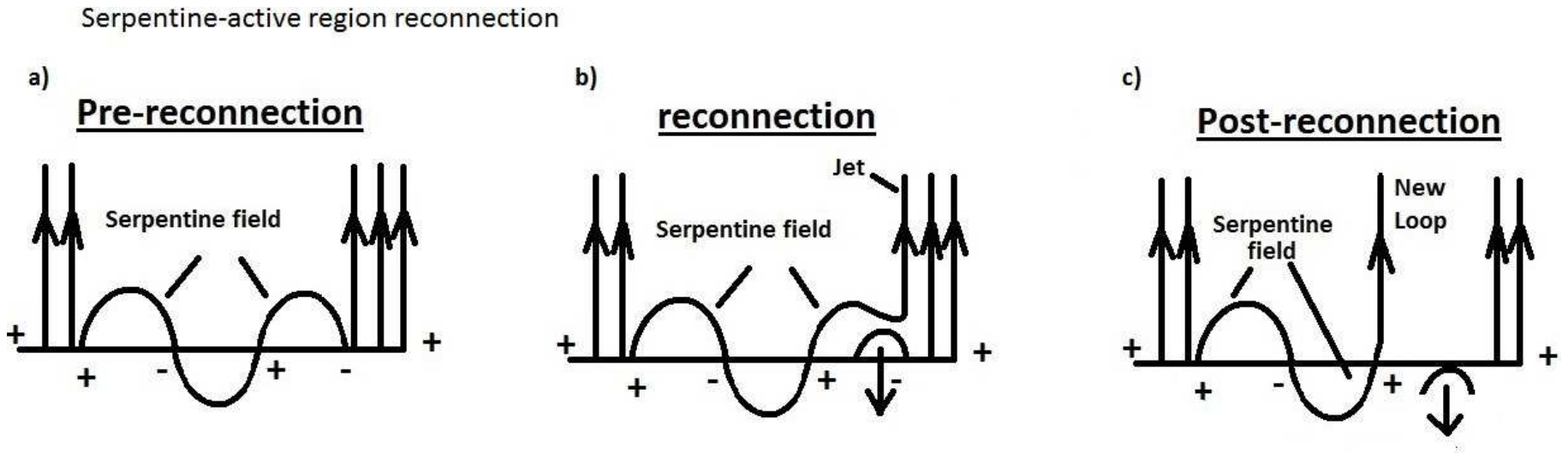}
\caption[Cartoon of early jet formation]{Cartoon showing the magnetic field layout of the active region and the serpentine field. (a) The magnetic configuration before reconnection occurs highlighting the serpentine field and the pre-existing field. (b) The reconnection occurs to form the jet and the smaller reconnection loop. (c) The new magnetic configuration after reconnection. This cartoon describes where and how the jets seen in Figure \ref{firstjets} were created.}
\label{jetearly}
\end{figure}

\begin{figure}[!ht]
\includegraphics[width=0.70\textwidth]{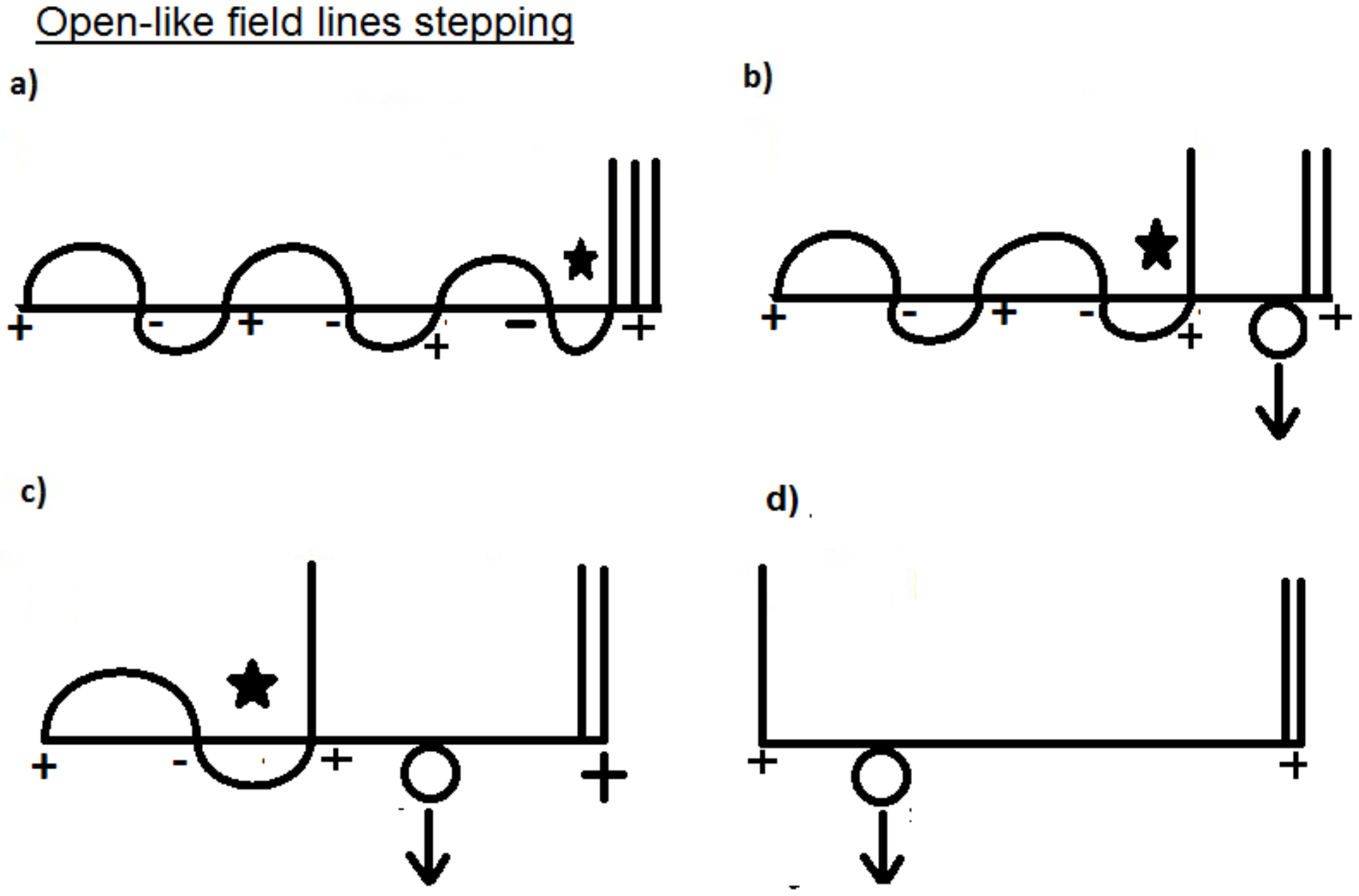}
\caption[Cartoon of serpentine field reconnection]{Cartoon showing the open-like field stepping over the serpentine field via interchange reconnection. The black star represents where reconnection can take place and the circle shows the reconnected field lines that cannot overcome magnetic tension.}
\label{serpentine}
\end{figure}

\indent The coronal jets seem to agree with what has been seen in previous work on jets in the solar atmosphere (\opencite{hpr77}; \opencite{shibataetal92}). Figure \ref{jetearly} describes the formation of the jets seen along the serpentine field. Magnetic reconnection occurs between the negative polarity of the serpentine field and the positive polarity of the active region (Figure\ref{jetearly}a). This is the favourable site for magnetic reconnection as we have oppositely directed \textbf{field lines}. Magnetic reconnection occurs at this location and a jet is released (Figure \ref{jetearly}b). Current reconnection models suggest that magnetic reconnection creates a small magnetic loop and a larger open-like field line that now has its footpoint in the positive field of the EFR. \textbf{This process is described by interchange reconnection (for example, \opencite{delzanna2011})}. In our case, the new foopoint location is associated with the positive polarity of the serpentine field (Figure \ref{jetearly}c). In Figure \ref{brightening4}, we see that the brightening is associated with the jet where magnetic reconnection is favourable within the serpentine field. As the serpentine continues to emerge, the jets (and the open-like field) in the chromosphere and corona migrate across the serpentine field, towards the north part of the EFR (Figure \ref{serpentine}). The appearance of coronal jets over the serpentine field means that the emerging serpentine field is able to release energy and contribute to the heating of the corona. This occurs once the serpentine field and the pre-existing active region field reconnect.

\begin{figure}[!ht]
\includegraphics[width=1.00\textwidth]{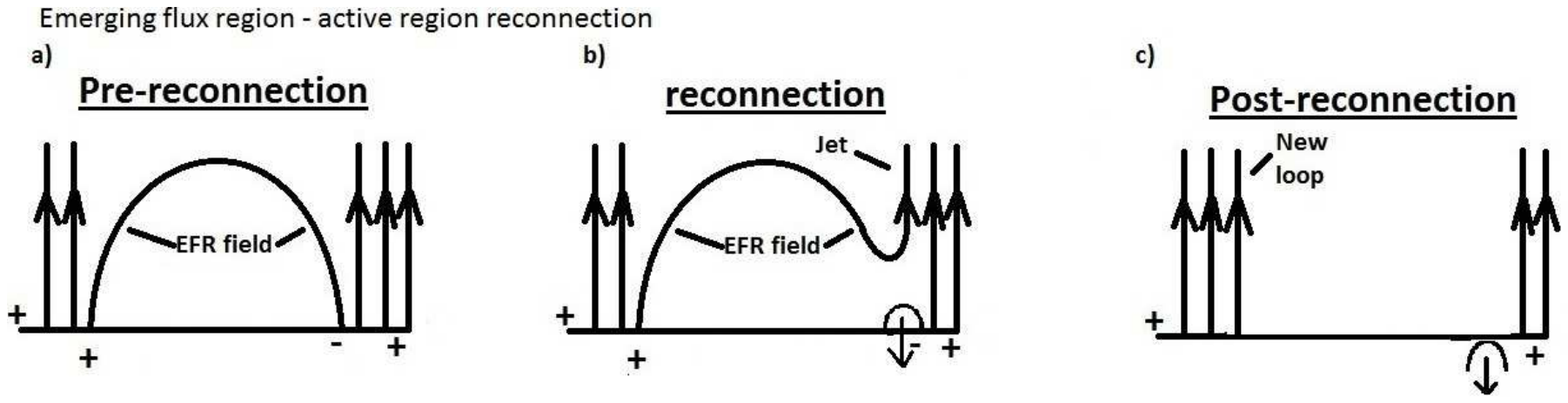}
\caption[Cartoon showing late jet formation]{Cartoon showing the magnetic field layout of the active region and the large-scale field of the EFR.  This includes the global EFR loop in panel (b) and the new loops highlighted on the left-hand side of panel (c). The cartoon describes how the jets seen in Figure 

\ref{jetlate}, (a) and (b) were created. The cartoon also describes how the intensity enhancements seen in the light curves in Figure \ref{AIAlightcurves} are connected to the emergence of the new flux.}
\label{jetlatecar}
\end{figure}

\subsection{The evolution of the large-scale EFR field}

\indent At later stages (Figure \ref{jetlatecar}) of the magnetic flux emergence (approximately eight hours after the magnetic flux emergence begins), the serpentine field has disappeared following successive reconnections, leaving two major polarities of the EFR. The coronal loops now form between these two polarities, but the negative polarity of the flux emergence is still interacting with the positive polarity of the active region (Figure \ref{jetlatecar}a). So therefore, magnetic reconnection continues to take place between the negative polarity of the EFR and the positive polarity of the active region (which is connected to a negative spot to the west). The pre-existing magnetic field lines are larger when compared to the field lines associated with the emerging flux. As before, a smaller new loop is formed at the reconnection site. However, the new larger loop that forms is now over the major positive polarity of the EFR and so the jets are formed along this loop (Figure \ref{jetlatecar}b). We see brightenings near the jet footpoint which is the favourable site for magnetic reconnection (Figure \ref{brightening5}). The jets at this time appear to follow a fan-like structure. Following successive reconnections between the negative polarity of the EFR and the positive polarity of the pre-existing active region, new loops start to build up on the north side of the EFR (Figure \ref{jetlatecar}c) as we see in the observations (Figure \ref{AIAlightcurves}). The blueshift enhancements seen in Figure \ref{EIShmi} are attributed to new upflows being created through magnetic reconnection between new flux and the pre-existing active region (the interaction between the negative flux of the EFR and the positive flux of the pre-existing active region). The footpoints of the new loops are associated with the upflow enhancements that we see in EIS. This suggests that chromospheric evaporation is occurring at the loop footpoints as a response to magnetic reconnection.

\section{Conclusion}
\label{Conclusion}
This study investigates how flux emergence affects a pre-existing active region.
EUV brightenings first form at the edge of the EFR and then form over the serpentine field approximately one hour later. The brightenings start approximately two hours after the flux emergence begins and ends at the same time as the flux emergence ends. The brightenings are seen between the photosphere and the corona and suggests that successive reconnections are required before brightenings can occur at the higher layers of the atmosphere.
It is possible for coronal jets to form over the serpentine field as the serpentine field is emerging. This has not been seen before and this is possibly due to the lack of high temporal and spatial resolution. This tells us that the emerging serpentine field can release the energy required to produce jets and heat the corona. The jets are also seen to migrate across the serpentine via interchange reconnection. After the serpentine field has fully emerged, jets are seen forming over the large-scale EFR field. We see that the jets and brightenings are co-spatial. However, the majority of non-jet related brightenings follow the serpentine field models of, e.g. \opencite{pasgrb04} or \opencite{Cheungetal08}.
Large-scale coronal loops were seen forming up to 43 Mm away from the EFR. The new loops form to the north of the EFR and the loop footpoints are linked to the upflow enhancements seen in EIS. These upflow enhancements have velocities of approximately 10 km \textbf{s$^{-1}$}. These are quite slow when compared to other upflow studies (e.g. \opencite{hmhtow10}) and suggests that different sized EFRs could create upflows with different velocities. However, some of the difference in velocity could be linked to the \textbf{orientation of the EFR with respect to the field line of the pre-existing active region (e.g. \opencite{Galsgaard2005})}. The link between the loop footpoints and the upflow enhancements suggest that chromospheric evaporation is occuring at the loop footpoints.

\begin{acks}
DLS would like to thank STFC for support via a studentship and LMG would like to thank the Royal Society for a fellowship.
Hinode is a Japanese mission developed and launched by ISAS/JAXA, collaborating with NAOJ as a domestic partner, NASA and STFC (UK) as international partners. Scientific operation of the Hinode mission is conducted by the Hinode science team organised at ISAS/JAXA. This team mainly consists of scientists from institutes in the partner countries. Support for the post-launch operation is provided by JAXA and NAOJ (Japan), STFC (UK), NASA, ESA, and NSC (Norway). SDO/AIA and SDO/HMI data are courtesy of NASA/SDO, the AIA and HMI science teams. 
\end{acks}

\bibliographystyle{spr-mp-sola-cnd}
\bibliography{thesisbib}

\end{article} 
\end{document}